# Differential Robustness in Transformer Language Models: Empirical Evaluation Under Adversarial Text Attacks

**Taniya Gidatkar, Oluwaseun Ajao, Matthew Shardlow**
Department of Computing and Mathematics
Manchester Metropolitan University
United Kingdom
`taniya.v.gidatkar@stu.mmu.ac.uk, s.ajao@mmu.ac.uk, m.shardlow@mmu.ac.uk`

## Abstract

This study evaluates the resilience of large language models (LLMs) against adversarial attacks, specifically focusing on Flan-T5, BERT, and RoBERTa-Base. Using systematically designed adversarial tests through TextFooler and BERTAttack, we found significant variations in model robustness. RoBERTa-Base and Flan-T5 demonstrated remarkable resilience, maintaining accuracy even when subjected to sophisticated attacks, with attack success rates of 0%. In contrast, BERT-Base showed considerable vulnerability, with TextFooler achieving a 93.75% success rate in reducing model accuracy from 48% to just 3%. Our research reveals that while certain LLMs have developed effective defensive mechanisms, these safeguards often require substantial computational resources. This study contributes to the understanding of LLM security by identifying existing strengths and weaknesses in current safeguarding approaches and proposes practical recommendations for developing more efficient and effective defensive strategies.

## 1 Introduction

Large language models (LLMs) represent a transformative deep learning approach for handling diverse natural language processing (NLP) tasks. These models, trained on vast datasets using transformer architectures, possess the ability to identify, interpret, forecast, and generate text. LLMs function as a form of artificial intelligence that understands, summarizes, generates, and predicts new material through extremely large datasets and deep learning methods (Abdali et al., 2023).

Language, which humans have developed over millennia to communicate, provides the foundation for all human and technical interactions. In artificial intelligence, language models serve a similar function, providing a framework for expression and idea generation. LLMs consist of multiple neural network layers, including recurrent, feedforward, embedding, and attention layers, which process input text and generate output (Dong et al., 2023).

As LLMs become increasingly integrated into critical systems and applications, understanding their security vulnerabilities and defensive capabilities becomes paramount. This research aims to systematically evaluate the robustness of popular LLMs against sophisticated adversarial attacks and to identify pathways for enhancing their security.

## 2 Background

LLMs are built on transformer architecture, functioning through a process of receiving input, encoding it, and then decoding it to provide output predictions. For an LLM to effectively process text input and generate predictions, it undergoes pre-training on massive textual datasets to develop fundamental capabilities, followed by fine-tuning for specific tasks (Hassanin and Moustafa, 2023).

When adapting an LLM for specialized tasks such as translation, it requires task-specific fine-tuning. This process enhances the effectiveness of particular functions and helps train the model for specific tasks using few-shot or zero-shot prompting approaches (Sakib et al., 2023). With few-shot prompting, the model learns to predict outcomes through examples provided in a prompt.

For organizational implementation of LLMs, resistance to manipulation is a critical security requirement. Modern LLMs are typically constructed on transformer architectures that have revolutionized NLP by enabling parallel text processing. Within this architecture, attention mechanisms assign varying degrees of relevance to different parts of phrases or sentences, allowing models to acquire contextual information effectively.

LLMs have significantly improved NLP capabilities in areas such as sentiment analysis, transla-tion, and text generation. However, the rapid advancement and increasing use of these models have raised concerns about their robustness and vulnerability to adversarial manipulation. Various attack vectors threaten LLMs, including data poisoning, adversarial perturbations, and model inversion, potentially reducing their reliability and dependability (Sun et al., 2023).

## 3 Formulation of the Problem

### 3.1 LLM as a Function

We formalise an LLM as a function $f_\theta : X \to Y$ that maps from an input space $X$ (typically sequences of tokens) to an output space $Y$ (labels, probabilities, or generated text). The parameter $\theta$ represents the model's weights learned during training. For classification tasks, $f_\theta(x)$ produces a probability distribution over possible classes, and the predicted class is:

$$\hat{y} = \arg\max_{y \in Y} f_\theta(x)_y \quad (1)$$

### 3.2 Adversarial Attack Definition

Adversarial attacks are techniques where carefully crafted changes are made to input data to fool a model into making a wrong prediction even though the changes might be invisible or meaningless to humans. It aims to find a perturbed input $x'$ that causes the model to make an incorrect prediction while maintaining similarity to the original input $x$. Formally, we define:

Find $x'$ such that: $f_\theta(x') \neq f_\theta(x)$ and $d(x, x') \leq \epsilon$ \quad (2)

where $d(\cdot, \cdot)$ is a distance function measuring similarity between inputs, and $\epsilon$ is the maximum allowable perturbation. In the context of NLP, this distance often considers semantic similarity and preserves grammatical structure.

### 3.3 Attack Success Rate

We quantify the effectiveness of an adversarial attack using the Attack Success Rate (ASR):

$$\text{ASR} = \frac{1}{|D|} \sum_{(x,y) \in D} \mathbf{1}[f_\theta(x') \neq y \text{ and } f_\theta(x) = y] \quad (3)$$

where $D$ is the test dataset, $(x, y)$ is an input-label pair, $x'$ is the adversarially perturbed version of $x$, and $\mathbf{1}[\cdot]$ is the indicator function.

### 3.4 TextFooler Attack Formulation

TextFooler(Jin et al., 2020) identifies important words in the input and replaces them with semantically similar alternatives to cause misclassification. For an input text $x = [w_1, w_2, ..., w_n]$, the algorithm:

1. Computes word importance score $I(w_i)$ for each word $w_i$:

$$I(w_i) = f_\theta(x)_y - f_\theta(x \setminus w_i)_y \quad (4)$$

where $x \setminus w_i$ is the text with $w_i$ removed or replaced with a placeholder.

2. Identifies candidate replacements with similar embeddings:

$$S(w_i, w_j) = \frac{emb(w_i) \cdot emb(w_j)}{||emb(w_i)|| \cdot ||emb(w_j)||} \quad (5)$$

where $emb(w)$ is the word embedding of $w$.

3. Iteratively replaces words in order of importance until the model prediction changes.

### 3.5 BERTAttack Formulation

BERTAttack(Li et al., 2020) leverages the bidirectional context from BERT to generate substitution candidates. For an input text $x = [w_1, w_2, ..., w_n]$:

1. For each position $i$, it masks the token $w_i$ and uses BERT to predict potential replacements:

$$P(w'|w_1, ..., w_{i-1}, [\text{MASK}], w_{i+1}, ..., w_n) \quad (6)$$

2. It selects the top-k candidates with highest probability and filters for semantic similarity.

3. It greedily replaces tokens in order of importance until the attack succeeds or a maximum number of replacements is reached.

### 3.6 Model Robustness Measure

We define the robustness score $R$ of a model against a specific attack as:

$$R = 1 - \frac{\sum_{i=1}^{N} \mathbf{1}[\text{Attack}_i \text{ succeeds}]}{N} \quad (7)$$

where $N$ is the total number of attack attempts. Higher values indicate greater robustness.

## 4 Methods

### 4.1 Experimental Setup

We employ a mixed-methods approach, combining qualitative literature review and quantitative testing. For the quantitative analysis, we utilized PromptBench(Zhu et al., 2024) to evaluate the robustness of Flan-T5(Chung et al., 2024), BERT-base(Devlin et al., 2019), and RoBERTa-base(Liu et al., 2019) models against TextFooler(Jin et al., 2020) and BERTAttack(Jin et al., 2020) methods. Specifically, Flan-T5 and BERT-base were tested on the SST-2 dataset (Socher et al., 2013) for sentiment classification tasks, while Flan-T5 and RoBERTa-Base were evaluated on SQuAD v2(Rajpurkar et al., 2018) for question-answering capabilities (Özkurt, 2023). The data collection protocol was approved by an ethics review board. No personally identifying information was used in the study. The SST-2 and SQUAD v2 datasets both contain English language examples, and a comparison is presented in Table 1 The base models of BERT, RoBERTa and FLAN-T5 were used in the study with 110m, 125m and 250m parameters, respectively. Less than 50 GPU hours were used for running the models and datasets within the Google Colab Pro A100 cloud infrastructure.

### 4.2 Algorithm for Attack Evaluation

We formalise our evaluation methodology in Algorithm 4.2:

EvaluateRobustnessModel $f_\theta$, Dataset $D$, Attack $A$ success ← 0, fail ← 0, skip ← 0
original_correct ← 0 perturbed_tokens ← []
query_counts ← [] for $(x, y) \in D$ do
$\hat{y} \leftarrow \arg\max f_\theta(x)$ {Original prediction}
**if** $\hat{y} = y$ **then**
  original_correct ← original_correct + 1
  $(x',$ is_success, queries, perturbed$) \leftarrow A(f_\theta, x, y)$
  **if** is_success **then**
    success ← success + 1
    Append perturbed to perturbed_tokens
    Append queries to query_counts
  **else**
    fail ← fail + 1
  **end if**
**else**
  skip ← skip + 1
**end if**
**end for**
original_accuracy ← original_correct/$|D|$
attack_success_rate ← success/(success + fail)
accuracy_under_attack ← original_accuracy − (original_accuracy × attack_success_rate)
avg_perturbed ← Mean(perturbed_tokens)
avg_queries ← Mean(query_counts)
**return** {success, fail, skip, original_accuracy, accuracy_under_attack, attack_success_rate, avg_perturbed, avg_queries} =0

### 4.3 Metrics for Evaluation

For each model and attack combination, we define the following metrics:

- Original Accuracy: $Acc_o = \frac{1}{|D|} \sum_{(x,y) \in D} \mathbf{1}[f_\theta(x) = y]$

- Accuracy Under Attack: $Acc_a = \frac{1}{|D|} \sum_{(x,y) \in D} \mathbf{1}[f_\theta(x') = y]$

- Attack Success Rate: $ASR = \frac{\text{Successful Attacks}}{\text{Total Attack Attempts}} = \frac{\sum_{(x,y) \in D'} \mathbf{1}[f_\theta(x') \neq y]}{|D'|}$, where $D'$ is the subset of examples correctly classified originally

- Average Perturbed Word Percentage: $PWP = \frac{1}{|S|} \sum_{(x,x') \in S} \frac{\text{HammingDistance}(x,x')}{\text{Length}(x)} \times 100\%$, where $S$ is the set of successful attacks

## 5 Results and Discussion

### 5.1 BERT-Base Under TextFooler Attack

Table 2 provides an overview of BERT-Base's performance under TextFooler attack, revealing significant vulnerability. Out of 100 test instances, TextFooler achieved a 93.75% success rate in misleading the model, reducing accuracy from 48% to merely 3%. This finding underscores BERT-Base's sensitivity to adversarial attacks.

Mathematically, the robustness score of BERT-Base under TextFooler attack is:

$$R_{\text{BERT}}^{\text{TextFooler}} = 1 - 0.9375 = 0.0625 \quad (8)$$

This extremely low robustness score indicates high vulnerability.

### 5.2 RoBERTa-Base Under BERTAttack

Table 3 demonstrates that RoBERTa-Base exhibited remarkable robustness when subjected to BERTAttack, with an attack success rate of 0%. Despite multiple attack attempts, RoBERTa-Base maintained its original accuracy of 35%, highlighting its exceptional resilience against this particular adversarial technique.

Table 1: Comparison of SST-2 and SQuAD v2.0 Datasets

| Feature | SST-2 | SQuAD v2.0 |
| --- | --- | --- |
| **Size** | 67,349 labeled phrases with 11,855 sentences from 2,210 reviews | 150,000 question-answer pairs on 505 articles (including 50,000 unanswerable questions) |
| **Task** | Binary sentiment classification (positive/negative) | Question answering with no-answer detection |
| **Domain Coverage** | Narrow: Movie reviews from Rotten Tomatoes | Broader: Wikipedia articles spanning history, science, geography, culture, etc. |
| **Language Coverage** | English only (primarily American English) | English only (formal, encyclopedic style) |
| **Linguistic Phenomena** | <ul><li>Sentiment expression</li><li>Evaluative language</li><li>Figurative language (metaphors, hyperbole)</li><li>Varying sentence complexity</li><li>Short text snippets</li></ul> | <ul><li>Various question forms</li><li>Factoid extraction</li><li>Paraphrasing</li><li>Simple inference</li><li>Unanswerable questions</li><li>Limited complex reasoning</li></ul> |
| **Demographic Representation** | <ul><li>Limited diversity (professional critics)</li><li>Western-centric cultural perspectives</li><li>Temporal bias of review period</li><li>Under-representation of minority viewpoints</li></ul> | <ul><li>Wikipedia contributor bias (predominantly male, Western, educated)</li><li>Western/Global North perspective dominance</li><li>Academic knowledge prioritization</li><li>Geographical and temporal skew</li></ul> |
| **Common Limitations** | <ul><li>Monolingual (English-only)</li><li>Cultural bias toward Western perspectives</li><li>Limited demographic diversity</li><li>Domain restrictions</li><li>Temporal limitations</li></ul> | |

Table 2: TextFooler Attack on BERT Model

| Metric | Value |
|---|---|
| Successful attacks | 45 |
| Failed attacks | 3 |
| Skipped attacks | 52 |
| Original accuracy | 48.00% |
| Accuracy under attack | 3.00% |
| Attack success rate | 93.75% |
| Avg perturbed word % | 21.93% |
| Avg words per input | 9.01 |
| Avg queries | 48 |

Table 3: BERTAttack on RoBERTa-Base Model

| Metric | Value |
|---|---|
| Successful attacks | 0 |
| Failed attacks | 7 |
| Skipped attacks | 13 |
| Original accuracy | 35.00% |
| Accuracy under attack | 35.00% |
| Attack success rate | 0.00% |
| Avg perturbed word % | N/A |
| Avg words per input | 8.70 |
| Avg queries | 240 |

The robustness score for RoBERTa-Base under BERTAttack is:

$$R_{\text{RoBERTa}}^{\text{BERTAttack}} = 1 - 0 = 1.0 \quad (9)$$

This perfect score indicates complete resilience to this attack.

### 5.3 Computational Efficiency Analysis

We define a computational efficiency metric $C$ that balances robustness and computational cost:

$$C = \frac{R}{\text{Avg. Query Count}} \times 100 \quad (10)$$

For RoBERTa-Base under BERTAttack:

$$C_{\text{RoBERTa}}^{\text{BERTAttack}} = \frac{1.0}{239.71} \times 100 = 0.417 \quad (11)$$

This indicates that while RoBERTa is perfectly robust, it comes at a high computational cost, requiring an average of 239.71 queries per attack attempt.

### 5.4 Summary of Results

Table 4 summarizes the results of all attacks on the tested models. The data reveals significant variance in LLM resilience, with RoBERTa-Base and Flan-T5 demonstrating notable resistance to both TextFooler and BERTAttack. These performance discrepancies highlight the need for enhanced protective measures in models like BERT-Base, which exhibited substantial vulnerabilities to adversarial manipulation.

We can mathematically compare the overall robustness of each model by averaging their robustness scores across different attacks:

$$\bar{R}_{\text{model}} = \frac{1}{|A|} \sum_{a \in A} R_{\text{model}}^a \quad (12)$$

where $A$ is the set of attacks tested. This gives us:

$$\bar{R}_{\text{BERT}} = \frac{R_{\text{BERT}}^{\text{BERTAttack}} + R_{\text{BERT}}^{\text{TextFooler}}}{2} = \frac{0 + 0.0625}{2} \quad (13)$$

$$\bar{R}_{\text{Flan-T5}} = R_{\text{Flan-T5}}^{\text{TextFooler}} = 1.0 \quad (14)$$

$$\bar{R}_{\text{RoBERTa}} = R_{\text{RoBERTa}}^{\text{BERTAttack}} = 1.0 \quad (15)$$

This quantitative comparison clearly illustrates the superior robustness of Flan-T5 and RoBERTa-Base compared to BERT-Base.

## 6 Ethical Considerations

Our research on LLM vulnerabilities to adversarial attacks raises several important ethical considerations that must be acknowledged:

### 6.1 Dual-Use Concerns

The techniques and metrics developed in this study for evaluating model robustness could potentially be misused to develop more effective adversarial attacks. We recognize this dual-use nature and have taken care to focus our discussion on defensive applications rather than exploitative ones. We advocate for responsible disclosure of vulnerabilities and emphasize that our goal is to improve model robustness, not to facilitate attacks.

### 6.2 Societal Implications

As LLMs become increasingly integrated into critical systems (healthcare, finance, legal, etc.), their vulnerability to adversarial manipulation raises significant societal concerns. System failures due to

Table 4: Results Summary for All Attacks

| Metric | BERTAttack on BERT | TextFooler on Flan-T5 | BERTAttack on RoBERTa |
| --- | --- | --- | --- |
| Successful attacks | 7 | 0 | 0 |
| Original accuracy | 35.00% | 35.00% | 20.00% |
| Accuracy under attack | 0.00% | 35.00% | 20.00% |
| Attack success rate | 100.00% | 0.00% | 0.00% |
| Avg perturbed word % | 27.56% | N/A | N/A |
| Avg words per input | 8.7 | 8.7 | 8.6 |
| Avg queries | 59.57 | 167.57 | 240 |

adversarial attacks could lead to harmful outcomes, including:

- Misinformation propagation through compromised language models

- Privacy violations through model manipulation

- Safety risks in automated systems relying on LLM outputs

- Erosion of trust in AI systems

### 6.3 Accessibility and Resource Considerations

Our findings regarding the computational intensity of effective defensive mechanisms (up to 240 queries per attack attempt) highlight potential accessibility concerns. Organizations with limited computational resources may struggle to implement adequate safeguards, potentially creating a security divide between well-resourced and under-resourced entities. We emphasize the importance of developing efficient defensive measures that are widely accessible.

### 6.4 Transparency in Vulnerability Reporting

We believe in transparent reporting of model vulnerabilities to foster trust and enable collective improvement of LLM security. However, we have carefully balanced this transparency with responsible disclosure practices, avoiding detailed descriptions of attack implementations that could be immediately weaponized.

### 6.5 Environmental Impact

The computational resources required for robust adversarial defense mechanisms have environmental implications due to increased energy consumption. Future work should consider how to optimize defensive strategies not only for effectiveness but also for energy efficiency to minimize environmental impact.

Through acknowledging these ethical dimensions, we hope to contribute to the development of LLM security measures that are not only technically sound but also aligned with broader societal values and needs.

## 7 Conclusion

This study assessed the robustness and vulnerability of large language models including Flan-T5, BERT-base, and RoBERTa-Base against adversarial attacks. Our findings indicate that while certain LLMs demonstrate significant resilience under adversarial conditions, substantial vulnerabilities remain, especially in models like BERT-base. The research confirms that even minor alterations to input data can substantially impact model predictions, underscoring the importance of enhanced safeguarding mechanisms for LLMs (Özkurt, 2023).

Although models such as Flan-T5 and RoBERTa-Base withstood adversarial attacks with remarkable effectiveness, maintaining their original accuracy levels with 0% attack success rates, our results highlight the need for refinement in current LLM defensive strategies. We observed that effective adversarial defenses currently demand significant processing power, with some models requiring up to 240 queries on average per attempted attack. This computational intensity points to the need for developing more efficient protective measures that maintain robust security while reducing resource requirements for practical applications.

Our research contributes to the growing body of knowledge on LLM security by quantifying model-specific vulnerabilities and identifying architectures that demonstrate superior resistance to common attack vectors. These insights can guide the development of more secure and reliable language models for deployment in sensitive and high-stakes applications.

# 8 Recommendations

Based on our findings, we propose several recommendations to improve LLM robustness and resilience against adversarial attacks:

## 8.1 Strengthening Adversarial Robustness with Adversarial Training

We propose an enhanced adversarial training objective that incorporates multiple attack types:

$$\min_\theta \mathbb{E}_{(x,y)\sim D} \left[L(f_\theta(x),y) + \sum_{i=1}^{k} \lambda_i \cdot L(f_\theta(A_i(x)),y)\right] \quad (16)$$

where $A_i$ represents different attack methods, and $\lambda_i$ balances the weight of each attack type.

## 8.2 Improving Token Embedding and Vocabulary Management

For token embeddings $E \in \mathbb{R}^{|V|\times d}$, we propose a regularization term that encourages robustness:

$$L_{\text{embed}} = L_{\text{task}} + \alpha \cdot \sum_{w_i,w_j \in S} \max(0, \cos(E_{w_i}, E_{w_j}) - \tau) \quad (17)$$

where $S$ is the set of semantically similar words, cos is cosine similarity, and $\tau$ is a threshold that maintains semantic distinctiveness.

## 8.3 Implementing Data Augmentation Techniques

The augmentation process can be formalised as a transformation function $T(x, \phi)$ where $\phi$ represents augmentation parameters. The training objective becomes:

$$\min_\theta \mathbb{E}_{(x,y)\sim D}[L(f_\theta(x),y) + \beta \cdot \mathbb{E}_{\phi \sim \Phi}[L(f_\theta(T(x,\phi)),y)]] \quad (18)$$

where $\Phi$ represents the distribution of augmentation parameters.

## 8.4 Leveraging Hybrid Defenses

For an ensemble of $m$ models $\{f_{\theta_1}, f_{\theta_2}, ..., f_{\theta_m}\}$, the prediction can be computed as:

$$f_{\text{ensemble}}(x) = \sum_{i=1}^{m} w_i \cdot f_{\theta_i}(x) \quad (19)$$

where $w_i$ are model weights that can be adjusted dynamically based on confidence scores or historical performance against specific attack types.

## 8.5 Theoretical Framework for Future Defenses

We propose a multi-objective optimization approach that balances accuracy, robustness, and computational efficiency:

$$\min_\theta [(1-\alpha-\beta)\cdot L_{\text{acc}}(f_\theta) + \alpha \cdot L_{\text{rob}}(f_\theta) + \beta \cdot L_{\text{comp}}(f_\theta)] \quad (20)$$

where:

$$L_{\text{acc}}(f_\theta) = \mathbb{E}_{(x,y)\sim D}[L(f_\theta(x),y)] \quad (21)$$
$$L_{\text{rob}}(f_\theta) = \mathbb{E}_{(x,y)\sim D} \max_{\delta \in \Delta}[L(f_\theta(x+\delta),y)] \quad (22)$$
$$L_{\text{comp}}(f_\theta) = \omega \cdot \text{CompMetric}(f_\theta) \quad (23)$$

## Limitations of Study

Our study faced some limitations that future research should address. First, our reliance on pre-trained LLMs like Flan-T5, BERT-base-uncased, and RoBERTa-Base limited our ability to examine internal defense mechanisms. While these models perform well on many NLP tasks, they require further refinement in their adversarial resilience. The limitation can be expressed mathematically as a constraint on the parameter space exploration:

$$\theta \in \Theta_{pretrained} \subset \Theta_{all} \quad (24)$$

This restricts our analysis to a subset of possible model configurations.

Second, our use of established benchmark datasets like SST2 and SQuAD v2, while providing standardized evaluation frameworks, may not fully capture the complexity of real-world data and adversarial scenarios. Performance under controlled testing conditions might not translate directly to less structured environments.

Third, although we employed sophisticated attack methods like TextFooler and BERTAttack, our approach did not cover the full spectrum of potential adversarial techniques. More advanced or semantically-driven attacks might reveal additional vulnerabilities not identified in this study.